\documentclass[amsmath, amsfonts, twocolumn, superscriptaddress,prx]{revtex4-1}
\usepackage{graphicx}
\usepackage{epsfig}
\usepackage{bm}
\usepackage{dcolumn}
\usepackage{amsmath}
\usepackage{amssymb}
\usepackage{color}
\usepackage[varg]{txfonts}

\newcommand{\rms}{\mathop{\rm rms}\nolimits}

\begin{document}

\title{Anomalous mesoscopic kinetics in disordered superconductors}

\author{Mengling Hettinger}
\affiliation{Department of Physics and Astronomy, Michigan State
University, East Lansing, Michigan 48824, USA}
\author{Maxim Khodas}
\affiliation{Racah Institute of Physics, Hebrew University of Jerusalem, Jerusalem 91904, Israel}
\author{Alex Levchenko}
\affiliation{Department of Physics, University of Wisconsin--Madison, Madison, Wisconsin 53706, USA}

\begin{abstract}
We study anomalous mesoscopic transport effects at the onset of the superconducting transition focusing on the observed large Nernst-Ettinghausen signal in disordered thin films. In the vicinity of the transition, as the Ginzburg-Landau coherence length of preformed Cooper pairs diverges, short-range mesoscopic fluctuations are equivalent to local fluctuations of the critical temperature. As a result, the dynamical susceptibility function of pair propagation acquires a singular mesoscopic component, and consequently, superconducting correlations give rise to enhanced mesoscopic fluctuations of thermodynamic and transport characteristics. In contrast to disordered normal metals, the root-mean-square value of mesoscopic conductivity fluctuations ceases to be universal and displays strong dependence on dimensionality, temperature, and under certain conditions can exceed its quantum normal state value by a large factor. Interestingly, we find different universality as magnetic susceptibility, conductivity, and transverse magnetic thermopower coefficients all display the same temperature dependence. Finally, we discuss an enhancement of mesoscopic effects in the Seebeck thermoelectricity and Hall conductivity fluctuations as mediated by emergent superconductivity.       
\end{abstract}

\date{May 2, 2019}

\maketitle

\section{Introduction}

Quantum-mechanical interference between different paths of electrons experiencing impurity scattering in conductors gives rise to important transport phenomena. The most notable examples are weak-localization and mesoscopic conductance fluctuations (see Ref. \cite{AA} for the review and references therein). Despite the fact that these effects are typically weak, $\delta\sigma/\sigma\sim 1/g\ll1$, where $\delta\sigma$ stands for either the weak-localization (WL) correction or root-mean-square (rms) value of conductance fluctuations, which are normalized to the Drude conductivity $\sigma$, and $g\gg1$ is the dimensionless conductance, they are fundamentally important. In particular, conductance fluctuations are universal at low temperatures and occur with the amplitude of conductance quantum $\delta\sigma\simeq e^2/(2\pi\hbar)$, where the exact numerical pre-factor depends only on whether time-reversal and/or spin-rotational symmetries are preserved. This universality persists as long as the characteristic sample size $L$ is smaller than dephasing length $L<L_\phi$. Furthermore, interaction effects in normal conductors barely change the magnitude and universality of conductance fluctuations (UCF), although they are crucially important in determining the temperature dependence of dephasing effects and, in particular, $L_\phi$ \cite{Aleiner-Blanter-PRB02}.

When superconductivity is induced at the boundary of the mesoscopic sample via the proximity effect, the universality of fluctuations remains almost intact \cite{Exp-UCF-SN-1,Exp-UCF-SN-2}. The only difference from the normal case is that the magnitude of oscillations changes by a number of the order of unity that can be traced to details of Andreev reflections at the superconductor-normal interface  \cite{BB-PRB95,Brouwer-PhD,HSL-PRB99}. Physics become quantitatively different if superconducting correlations are present in the bulk of the sample. Experimentally, this is achieved by tuning superconducting systems to the vicinity of the critical temperature $T_c$ or, alternatively, in the proximity (or across) of the superconductor-insulator transition. Compelling evidence, ranging from measurements in two-dimensional granular arrays \cite{Frydman-SSC98,Frydman-PU98}, sub-micron scale superconducting cylinders \cite{Zadorozhny-PRB01}, and quantum wires \cite{Dynes-PRB98,Shahar-PRL05}, exists that mesoscopic oscillations could become giant, sometimes reaching the level of $\sim 10^4\times e^2/(2\pi\hbar)$. These observations seemingly imply that the role of mesoscopic effects proliferates in the presence of superconducting correlations. 

It should be noted, however, that even without superconductivity there are circumstances when mesoscopic fluctuations become anomalously large.  One example is given by the Seebeck thermopower \cite{Ma}, and connected to it by the Onsager relation Peltier coefficient \cite{DiVincenzo},  another is Coulomb drag transresistance \cite{Narozhny-Aleiner,ALAK}. Indeed, in contrast to conductance, Seebeck coefficient fluctuations $\delta\alpha$ acquire an additional large factor, $\delta\alpha/\alpha\simeq (E_F/E_{\mathrm{Th}})(\delta\sigma/\sigma)$, in the ratio between Fermi and Thouless energies $E_F/E_{\mathrm{Th}}\gg1$. This happens because thermopower relies on particle-hole asymmetry, so its sample average value scales inversely proportionally to Fermi energy, which is in accordance with the Mott formula, $\alpha\propto \partial\sigma/\partial E_F$. On the other hand, at the mesoscopic level particle-hole symmetry is broken much more strongly on the scale of Thouless energy $E_{\mathrm{Th}}$, thus giving a substantial enhancement. For drag $\sigma_D$, particle-hole symmetry should be broken for both layers, so that enhancement of mesoscopic fluctuations in transconductance is even bigger in that parameter, namely, $\delta\sigma_D/\sigma_D\simeq (E_F/E_{\mathrm{Th}})^2\gg1$. Another crucial difference between these two examples is that drag and its variations occur solely due to interactions, whereas conductance and thermopower fluctuations are understood at the single-particle level.      

Returning to the discussion of the superconducting systems, we should mention that theoretical studies devoted to various aspects of mesoscopic fluctuations cover a diverse range of topics. These works include mesoscopic effects on thermodynamic properties such as the Josephson current \cite{AltshulerSpivak,SpivakZyuzin,Beenakker,Micklitz,HouzetSkvortsov,HXAL}, upper critical field \cite{SpivakZhou,GalitskiLarkin}, critical temperature \cite{SkvortsovFeigelman}, condensation energy and glassy phase transitions \cite{ZhouSpivak,Lamacraft}, density of states, gap fluctuations, and level statistics \cite{Melsen,AltlandSimonsSemchuk,Vavilov,Ostrovsky,Meyer,Kopnin,Burmistrov}; as well as some transport characteristics such as persistent and thermoelectric currents, and fluctuation conductivity \cite{Oppen,Buzdin,Schwiete,Wohlman,SpivakZyuzin-EPL98,ZhouBiagini}. Our main motivation is to investigate mesoscopic effects in thermomagnetic properties of disordered superconducting thin films. This research is primarily inspired by the measurements of the Nernst-Ettinghausen effect and diamagnetic response in superconductors \cite{Nernst-Exp1,Nernst-Exp2}, which revealed anomalously large signals, including high-$T_c$ \cite{Nernst-Exp3,Nernst-Exp4,Nernst-Exp5,Nernst-Exp6,Nernst-Exp7,Nernst-Exp8,Nernst-Exp9,Nernst-Exp10,Nernst-Exp11}, and heavy-fermion systems \cite{Nernst-Exp12,Nernst-Exp13,Nernst-Exp14}. The problem of finding a possible microscopic mechanism of the large Nernst effect attracted tremendous attention and triggered a number of theoretical proposals \cite{N-Th-Ussishkin-1,Kontani,N-Th-Ussishkin-2,N-Th-Oganesyan,N-Th-Hartnoll,Yakovenko,N-Th-Podolsky,N-Th-Raghu,N-Th-Reizer,N-Th-Serbyn,N-Th-Michaeli,N-Th-Ivar,Lidmar,N-Th-AL,Orgad,Fujimoto}. Most of these results, including experimental findings from multiple groups and theoretical approaches to address the data, were summarized in a recent review \cite{Nernst-Review}. 

In this work we show that the interplay of interactions in the Cooper channel and local mesoscopic fluctuations has a profound effect on kinetics of superconductors near $T_c$. In particular, we find that in the  temperature range of the Ginzburg region, $\mathrm{Gi}\ll (T-T_c)/T_c\ll 1$, with $\mathrm{Gi}\simeq 1/g$ being the Ginzburg number, where superconducting correlations manifest in the fluctuation-induced transport \cite{Book}, the temperature dependence of the variance in the transverse thermomagnetic response is strongly enhanced.     

The rest of the paper is organized as follows. In Sec. \ref{sec:qualitative} we provide qualitative arguments and estimates for the physical picture of strong mesoscopic fluctuations in superconductors. In Sec. \ref{sec:kinetics} we place these ideas on the firm footing of microscopic diagrammatic analysis. Specifically, we compute diagonal and Hall conductivities, longitudinal thermoelectric and transverse thermomagnetic coefficients, as well as magnetic susceptibility fluctuations. We close this paper in Sec. \ref{sec:discussion} with a brief discussion of the experimental situation and present ideas for further developments in the regime of quantum fluctuations.       

\section{Qualitative considerations} \label{sec:qualitative}

It was emphasized early on~\cite{SpivakZhou,ZhouSpivak} that quantum interference mesoscopic effects may lead to the formation of superconducting droplets that nucleate prior to transition of the whole system. Above $T_c$ there are also thermally induced superconducting fluctuations~\cite{Book} that are known to be crucially important in describing transport properties. One thus expects that the combined effect of two fluctuation mechanisms may have interesting implications for the kinetic properties of superconductors. Indeed, the probability amplitude of the fluctuations in the pairing gap is controlled by the competition of Cooper pair condensation energy and entropy and can be estimated from the Ginzburg-Landau (GL) functional. The condensation energy exhibits mesoscopic fluctuations with an amplitude $\propto 1/g$ and a correlation radius of the order of the thermal length $\sim L_T=\sqrt{D/T}$. Near $T_c$ the latter coincides with the superconducting coherence length $\xi=\sqrt{D/T_c}$, where $D$ is the diffusion coefficient of the metal in the normal state. On the other hand, thermal superconducting fluctuations are susceptible to Ginzburg-Landau correlation length $\xi_{\mathrm{GL}}=\xi\sqrt{T_c/(T-T_c)}\gg L_T$, so mesoscopic fluctuations are almost local with respect to superconducting fluctuations, and thus should be summed randomly from different blocks of size $\xi$. For the $d$-dimensional sample the number of such segments is $\sim(L/\xi)^{d}$. Let us denote by $\delta\Delta_{\xi}$ the random local fluctuation of the gap occurring on the scale of $\xi$, while we denote by $\delta\Delta_L$ its sample average value. The two are related as $\delta\Delta_L\sim(L/\xi)^{d/2}\delta\Delta_\xi$. Knowing $\delta\Delta_L$, one can estimate critical temperature
fluctuations $\delta T_c$ as follows $\delta T_c\sim\langle\delta\Delta^2_L\rangle/T_c$. 
This argument essentially comes from approximating the quartic
term in the GL functional
$\delta\Delta^4_L\approx\delta\Delta^2_L\langle\delta\Delta^2_L
\rangle$ and reabsorbing that term into the redefinition of $T_c$. Our goal is to estimate $\delta\Delta_\xi\to\delta\Delta_L\to\delta T_c$. The probability of fluctuation, 
$\mathcal{P}[\delta\Delta_\xi]\propto\exp[-
g(L/\xi)^2(\delta\Delta^2_\xi/T^2_c)]$,
can be deduced from the optimization of the GL functional.  
Indeed, the exponential factor comes from the gradient term in the GL action that governs spatial fluctuations of superconducting droplets: $\nu\int d^dr
D(\partial_r\delta\Delta_\xi)^2\sim\nu D
(L^d/\xi^2)\delta\Delta^2_\xi=\nu D L^{d-2}
(L/\xi)^2\delta\Delta^2_\xi$, where $g=\nu D L^{d-2}$ is the
dimensionless conductance of the $d$-dimensional cube. With this probability density one estimates the typical local fluctuation of the gap and, consequently, critical temperature 
$ \delta T_c/T_c\sim \mathrm{Gi}\, (\xi/L)^{(4-d)/2}$. These estimates and line of reasoning closely follow earlier ideas by Ioffe and Larkin, who considered superconductors with local fluctuations of $T_c$ within the phenomenological approach \cite{Ioffe-Larkin}. Ultimately, the dynamic pair susceptibility propagator, $P(\omega,q)\propto (Dq^2+T-T_c+|\omega|)^{-1}$, defined for a given mode at finite frequency $\omega$ and wave vector $q$, acquires an anomalous mesoscopic component $\delta P\propto P^2\delta T_c$. Even though the whole effect is small, $\delta T_c/T_c\ll1$, as it scales inversely proportionally to conductance, $g\gg1$, the singular nature of $P(\omega,q)$ at $T-T_c\ll T_c$ as $\{q,\omega\}\to0$ translates into the substantial temperature dependence of kinetic coefficients. This is the microscopic reason for the breakdown of the universality of mesoscopic effects in the case of fluctuating superconductors. Next, we elaborate these considerations within the microscopic diagrammatic formalism (throughout the text we use units of $\hbar= k_B=1$).

\section{Anomalous mesoscopic fluctuations in disordered superconductors} \label{sec:kinetics} 

\subsection{Definitions and assumptions}

We begin with the definition of kinetic coefficients concentrating on the linear response analysis. The electric $\bm{J}^e_{\mathrm{tr}}$ and heat $\bm{J}^h_{\mathrm{tr}}$ transport currents are related to the electric field $\bm{E}$ and temperature gradient $\bm{\nabla} T$ by the matrix of thermoelectric coefficients 
\begin{eqnarray}\label{eq:Je-Jh}
\left( \begin{array}{cc}\bm{J}^{e}_{\mathrm{tr}} \\ \bm{J}^{h}_{\mathrm{tr}} \end{array} \right) = \left( \begin{array}{cc} \hat{\sigma} & \hat{\alpha} \\ \hat{\beta} & \hat{\kappa} \end{array} \right)\left( \begin{array}{cc} \bm{E} \\ -\bm{\nabla} T \end{array} \right),
\end{eqnarray}
where $\hat{\sigma}$ is the electric conductivity tensor, $\hat{\alpha}$ and $\hat{\beta}$ are the thermoelectric tensors ($\hat{\beta} = T \hat{\alpha}$ due to Onsager relations), and $\hat{\kappa}$ is the thermal conductivity tensor. Applying the open-circuit condition to Eq. \eqref{eq:Je-Jh}, the Nernst coefficient is expressed in terms of the components of conductivity and thermoelectric tensors as follows:
\begin{equation}\label{eq:Nernst-Def}
N=\frac{E_y}{(-\nabla_xT)H}=\frac{1}{H}\frac{\alpha_{xy}\sigma_{xx}-\alpha_{xx}\sigma_{xy}}{\sigma^2_{xx}+\sigma^2_{xy}}.
\end{equation}
We assume that magnetic field $H$ is applied in the $z$-direction. 

The are two important aspects in the calculation of $N$ that need to be discussed. The first point concerns the role of magnetization currents. 
At the technical level of the Kubo formula, the microscopic electric $(\bm{J}^e)$ and heat $(\bm{J}^h)$ currents contain both transport and magnetization contributions, namely:  
$\bm{J}^e=\bm{J}^e_{\mathrm{tr}}+\bm{J}^e_{\mathrm{mag}}$, $\bm{J}^h=\bm{J}^h_{\mathrm{tr}}+\bm{J}^h_{\mathrm{mag}}$. In the presence of an applied electric field, it was shown in Ref.~~\cite{Cooper-Halperin-Ruzin} that the magnetization current is given by $\bm{J}^h_{\mathrm{mag}}=c\bm{M}\times\bm{E}$, where $\bm{M}$ is the equilibrium magnetization (in the absence of the electric field). Since the magnetization currents circulate in the sample they do not contribute to the net currents which are measured in a transport experiment. For that reason, the computation of $\alpha_{xy}$ comprises two independent derivations. In the first step, one finds the response of the total current to the applied electric and magnetic fields, and in the second step one finds the magnetization currents that should be derived from the equilibrium magnetization. It then follows that the transverse thermomagnetic response is given by subtracting these two contributions 
\begin{equation}
\alpha_{xy}=-\frac{J^h}{E_xT}+\frac{cM_z}{T}=\beta_{xy}+\frac{cM_z}{T}.
\end{equation}
Therefore, we need to know the magnetic susceptibility $\bm{M}=\chi\bm{H}$, which will be computed diagrammatically along with $\beta_{xy}$. The importance of the magnetization contribution to $\alpha_{xy}$ in the context of superconducting fluctuations was elaborated by Ussishkin~\cite{N-Th-Ussishkin-2}. 

The second point concerns the role of particle-hole asymmetry in response tensors. In the normal state, diagonal elements of electrical $\hat{\sigma}$ and thermal $\hat{\kappa}$ conductivities are present already at the level of perfect particle-hole symmetry, i.e., neglecting any contributions which arise due to asymmetry around the Fermi surface in properties such as the density of states and transport scattering time, whereas off-diagonal elements vanish. The conventional result for the thermoelectric tensor $\hat{\alpha}$ in the normal metal (the so-called quasiparticle contribution) also vanishes in this limit, as can be seen from the Mott formula for $\alpha_{xx}$ and Sondheimer formula for $\alpha_{xy}$. However, it was emphasized in Ref.~\cite{N-Th-Ussishkin-2} that this result is not required by the symmetry and will not necessarily hold when additional scattering processes, such as superconducting or mesoscopic fluctuations, are taken into account without breaking the particle-hole symmetry. In particular, accounting for superconducting fluctuations gives finite $\alpha_{xy}$ but still vanishing $\alpha_{xx}$ and $\sigma_{xy}$ without particle-hole asymmetry. In this case, the general expression for the Nernst coefficient Eq.~\eqref{eq:Nernst-Def} reduces to $N=\alpha_{xy}/(H\sigma_{xx})$. 

It is known that superconducting fluctuations enhance the conductivity close to $T_c$ due to the so-called Aslamazov-Larkin \cite{Aslamazov-Larkin} and Maki-Thompson \cite{Maki-Thompson} contributions as well as density of states effects \cite{Abrahams-DoS}. A similar identification of the microscopic contributions applies to other transport coefficients. In the case of the transverse thermomagnetic response, the leading-order contribution to $\alpha_{xy}$ is due to the Aslamazov-Larkin (AL) diagrams alone. The contributions of the Maki-Thompson (MT) and density of states (DOS) diagrams are less divergent as $T\to T_c$. This is true as long as we are discussing fluctuations in a weak field $H\ll H_{c2}$ near $T_c$. This picture is more complicated in the quantum critical regime $H\sim H_{c2}$ and $T\to0$ as all the terms happen to be of the same order. To further simplify our analysis we will assume $s$-wave symmetry of the superconducting order parameter. In the context of the high-temperature superconductors, it is of interest to consider also the case of $d$-wave symmetry in a similar approach. 

\subsection{Kubo formulas}

Within the linear response analysis, the diagonal Aslamazov-Larkin conductivity is determined from the following current-current Kubo kernel $K^{ee}_{xx}$ \cite{Aslamazov-Larkin}
\begin{align}\label{eq:K-xx}
&\sigma_{xx}=\lim_{\Omega\to0}\frac{1}{\Omega}\Im [K^{ee}_{xx}(\Omega)]^R, \nonumber \\
&K^{ee}_{xx}(\Omega_m)=4e^2T\sum_{q\omega_n}B^2_x(q)P(\omega_n,q)P(\omega_n+\Omega_m,q)\,,
\end{align}
where $[K^{ee}_{xx}]^R$ indicates the retarded component of $K^{ee}_{xx}(\Omega_m)$ as it is analytically continued from the discrete Matsubara frequencies into the entire complex plane $\mathrm{i}\Omega_m\to \Omega+\mathrm{i}0$. The pair susceptibility propagator of fluctuating Cooper pairs is of the form 
\begin{equation}\label{eq:P}
P(\omega,q)=-\frac{1}{\nu}\frac{1}{\ln(T/T_c)+\pi Dq^2/8T+\pi|\omega|/8T}
\end{equation}
which is an asymptotic formula valid at small momenta and frequencies, namely, $\{Dq^2,\omega\}\ll T$ (here $\nu$ is the single-particle density of states in the normal state). With the same accuracy one can treat $\ln(T/T_c)\approx (T-T_c)/T_c$. In the definition of the current response kernel we also made use of the current vertex 
\begin{equation}\label{eq:B-e}
B^e_i=2eB_i,\quad B_i(q)=-2\nu\eta q_i,\quad \eta=\pi D/8T
\end{equation}
which diagrammatically corresponds to the triangular block of electronic Green's functions of the AL diagram. This expression for $B^e(\omega,q)$ is derived under the same approximations for typical frequencies $\omega\sim (T-T_c)\ll T$ and momenta $q\sim \xi^{-1}_{\mathrm{GL}}\ll L^{-1}_T$ of superconducting fluctuations as in Eq. \eqref{eq:P}. As alluded to above, $B^e_i$ has a much more complicated structure in the regime of quantum fluctuations, and its frequency dependence plays a crucial role \cite{N-Th-Serbyn}. 

The corresponding Aslamazov-Larkin contribution to the transverse thermoelectric coefficient can be found from the
mixed electric current-heat current Kubo response function $K^{eh}_{xy}$ \cite{N-Th-Ussishkin-2}: 
\begin{equation}\label{eq:beta}
\beta_{xy}=\frac{H}{cT}\lim_{\Omega,Q\to0}\frac{1}{\Omega
Q}\Re[K^{eh}_{xy}(Q,\Omega_m)]^R,
\end{equation}
where
\begin{align}\label{eq:K-xy}
&K^{eh}_{xy}(\Omega_m,Q)=-4e^2T\sum_{q,\omega_n}B_x(q)B^2_y(q)(\mathrm{i}\omega_n+\mathrm{i}\Omega_m/2)\nonumber \\ &\times\left[P(\omega_n,q-Q_x)P(\omega_n,q)P(\omega_n+\Omega_m,q)\right. \nonumber\\  
&\left.+P(\omega_n,q)P(\omega_n+\Omega_m,q)P(\omega_n+\Omega_m,q+Q_x)\right]
\end{align}
with the heat vertex 
\begin{equation}\label{eq:B-h}
B^h_i(\omega_n,q)=2\mathrm{i}\nu\omega_n\eta q_i=-\mathrm{i}\omega_n B_i(q). 
\end{equation}

We finally define magnetic susceptibility from the equilibrium magnetization.  Diagrammatically, it can be calculated to linear order in $H$ by considering the current response to a magnetic field at a finite wave-vector $Q$ \cite{Aslamazov-Larkin-chi}:
\begin{equation}\label{eq:chi}
\chi_{\mu\nu}=-\frac{4e^2}{c^2}\epsilon_{\alpha\gamma\mu}\epsilon_{\beta\kappa\nu}T\sum_{\omega,q}\hat{x}_\gamma\hat{x}_\kappa P^2(\omega,q)\Pi'_\alpha(\omega,q)\Pi'_\beta(\omega,q),
\end{equation}
where $\epsilon_{\alpha\beta\gamma}$ is the anti-symmetric Levi-Civita unity tensor, $\hat{x}$ is the coordinate operator in the momentum representation, and $\Pi(\omega,q)$ is the electronic polarization operator given the usual loop diagram composed of the product of two Green's functions (we recall that resummation of these loops gives exactly Eq. \eqref{eq:P}). Below we will consider only the isotropic case $\chi_{\mu\nu}=\chi\delta_{\mu\nu}$. 

\subsection{Mesoscopic conductivity fluctuations}

With these technical prerequisites, we proceed with the calculation of superconductivity-induced mesoscopic fluctuations in $\sigma_{xx}$, $\beta_{xy}$, and $\chi$. In particular, we will compute their root-mean-square values, e.g. $\rms \{\sigma,\alpha,\chi\}$. The first step in the derivation of the defined kinetic coefficients requires the consideration of discrete sums over Matsubara frequencies $\omega_n=2\pi nT$. Such summations over bosonic frequencies can be conveniently done with the help of closed contour integration in the complex plane by using the following formula:
\begin{equation}\label{eq:f}
T\sum_{\omega_m} f(\omega_m)=\frac{1}{4\pi\mathrm{i}}\oint d\omega f(-\mathrm{i}\omega)\coth\left(\frac{\omega}{2T}\right).
\end{equation} 
In application of Eq. \eqref{eq:f} to Eq. \eqref{eq:K-xx} one notices that the product of two propagators under the integral has breaks of analyticity in the complex plane of $\omega$ at $\Im\omega=0$ and $\Im\omega=-\Omega_m$, so that the integration contour has two branch cuts along these lines. Following the standard steps of analytic continuation \cite{Book}, one arrives in the intermediate step at:  
\begin{equation}
\sigma_{xx}=\frac{e^2}{\pi T}\sum_qB^2_x(q)\int
\frac{[\Im P^R(\omega,q)]^2d\omega}{\sinh^2(\omega/2T)}.
\end{equation}  
Integrating over $q$ and $\omega$ with the help of Eqs. \eqref{eq:P} and \eqref{eq:B-e} one finds the celebrated Aslamazov-Larkin formula $\sigma_{xx}=(e^2/16)\ln^{-1}(T/T_c)$. Being interested in its mesoscopic fluctuations, we square this diagram and average it over the disorder potential, which gives
\begin{align}\label{eq:sigma-MF}
&\langle\delta\sigma^{2}_{xx}\rangle=\frac{4e^4}{\pi^2T^2}\sum_{q_1q_2}B^2_x(q_1)B^2_x(q_2)\nonumber \\ 
&\int\frac{M_{12}(\omega,q)
\Im P^R(\omega_1,q_1)\Im P^R(\omega_2,q_2)
d\omega_1
d\omega_2}{\sinh^2(\omega_1/2T)\sinh^2(\omega_2/2T)}.
\end{align}
In order to calculate the mesoscopic (disorder-irreducible) correlation function 
\begin{align}
M_{12}(\omega,q)=\langle\Im\delta P^R(\omega_1,q_1)\Im\delta P^R(\omega_2,q_2)\rangle
\end{align}
of the pairing susceptibility propagators, one has to draw two copies of diagrams for $P$, each representing a given realization of the disorder potential, and then connect their diffusive parts by additional impurity lines. Such construction involves four colliding diffuson-cooperon ladders and, on a technical level, requires computation of four- and six-order Hikami boxes \cite{Hikami}. Some of these diagrams have been studied before \cite{SpivakZhou,ZhouSpivak,GalitskiLarkin,SkvortsovFeigelman,Lamacraft,ZhouBiagini,Kee-Aleiner,Galitski-DasSarma,AL} and we invoke that knowledge for our purposes. In particular, the most singular contribution has the form  
\begin{align}
&\langle \delta P^{R(A)}(\omega_1,q_1)\delta P^{R(A)}(\omega_2,q_2)\rangle=\nonumber \\ 
&\frac{A\nu^2}{g^2}
\left(\frac{L_T}{L}\right)^2[P^{R(A)}(\omega_1,q_1)]^2[P^{R(A)}(\omega_2,q_2)]^2.
\end{align}
As anticipated [see discussion in Sec. \ref{sec:qualitative}] the induced mesoscopic effect is weak, $\langle \delta P^2\rangle\propto \mathrm{Gi}^2$, however it exhibits an extremely singular behavior in the long wave-length limit $\{\omega,q\}\to0$ where $\langle\delta P^2\rangle\propto (T-T_c)^{-4}$. The precise value of the numerical factor $A\sim1$ is not of principal importance in view of the strong dependence of the whole expression on temperature and system size. It is then straightforward to show that 
\begin{align}\label{eq:H}
&M_{12}=\frac{4A\nu^2}{g^2}\left(\frac{L_T}{L}\right)^2\times
\nonumber \\ 
&\Im P^R(\omega_1,q_1)
\Re P^R(\omega_1,q_1)
\Im P^R(\omega_2,q_2)
\Re P^R(\omega_2,q_2).
\end{align}  
For convenience in further integration, we define the following dimensionless units: $x=\eta q^2$, $y=\pi\omega/8T$, and $\epsilon=\ln(T/T_c)\approx(T-T_c)/T_c$. In these variables, the interaction propagator and vertex function become
\begin{equation}\label{eq:P-B-xy}
\Im P^{R}(x,y)=-\frac{1}{\nu}\frac{y}{(\epsilon+x)^2+y^2}\,,\quad
\mathcal{B}^2_x(x)=4\nu^2\eta x\cos^2\phi\,,
\end{equation}
and for a two-dimensional geometry of a thin superconducting film, integrations transform as
\begin{equation}
\sum_{q}\!\to\!\int^{2\pi}_{0}\!\!\int^{\infty}_{0}\!\!\frac{d\phi dx}{8\pi^2\eta}\,,
\int\frac{d\omega}{\sinh^2(\omega/2T)}\to\frac{\pi T}{2}\int^{+\infty}_{-\infty}\frac{dy}{y^2}\,,
\end{equation}
where we expanded $\sinh y\approx y$ since the major contribution comes from the range of parameters $\{x,y\}\sim\epsilon\ll1$. With these notations Eq. \eqref{eq:sigma-MF} becomes 
\begin{align}
&\langle\delta\sigma^{2}_{xx}\rangle=\frac{Ae^4L^2_T}{\pi^4g^2L^2}
\int^{2\pi}_{0}d\phi_1d\phi_2\cos^2\phi_1\cos^2\phi_2\nonumber\\ 
&\int^{\infty}_{0}
\int^{+\infty}_{-\infty}\frac{x_1x_2(x_1+\epsilon)(x_2+\epsilon)dx_1dx_2dy_1dy_2}
{[(x_1+\epsilon)^2+y^2_1]^3[(x_2+\epsilon)^2+y^2_2]^3}.
\end{align}
The integrations are elementary and can be made separable in all variables by rescaling first $y_i\to (x_i+\epsilon)y_i$ and then $x_i\to \epsilon x_i$. As a result, the root-mean-square value of conductivity fluctuations takes the form (suppressing the numerical factor of the order of unity)   
\begin{equation}\label{eq:rms-sigma-xx}
\rms \sigma_{xx}\simeq \sigma_Q \mathrm{Gi}\frac{L_T}{L}\left(\frac{T_c}{T-T_c}\right)^{2}    
\end{equation}
where we introduced the quantum of conductance $\sigma_Q=e^2/(2\pi\hbar)$. This estimate is valid at length scales $L>\xi_{\mathrm{GL}}$, whereas fluctuations saturate to $\rms \sigma_{xx}\sim \sigma_Q \mathrm{Gi}(T/E_{\mathrm{Th}})^{3/2}$ when  $\mathrm{min}\{L_T,\xi\}<L<\xi_{\mathrm{GL}}$. This happens because the continuous spectrum of soft superconducting excitations becomes quenched by the finite-size quantization and Thouless energy provides a natural cutoff, $Dq^2\to E_{\mathrm{Th}}$, in a pair-propagator.   

For completeness, we have also analyzed the mesoscopic Maki-Thompson part in fluctuation-induced diagonal conductivity, in particular its anomalous piece, which is the most singular near $T_c$. Technically it follows from the same current-current kernel $K^{ee}_{xx}$, but is given by a different diagram \cite{Maki-Thompson}. The corresponding analytically-continued expression is well known: 
\begin{align}
&\sigma^{\mathrm{an}}_{xx}=\frac{e^2\nu D}{2\pi T}\sum_q\int\frac{\coth(\omega/2T)d\varepsilon d\omega}{\cosh^2(\varepsilon/2T)}\nonumber \\
&\Im P^R(\omega,q)C^R(2\varepsilon+\omega,q)
C^A(2\varepsilon+\omega,q)
\end{align}
where 
\begin{equation}
C^R(\varepsilon,q)=\frac{1}{Dq^2+\tau^{-1}_\phi-\mathrm{i}\varepsilon}    
\end{equation}
is the cooperon (the summed impurity ladder in the particle-particle channel), and $\tau_\phi$ is its dephasing time. It should be noted that the same dephasing time should appear in the pair propagator as well, but its net effect is to shift the critical temperature $T_c\to T_c-\pi/(8\tau_\phi)$. An additional integral over $\varepsilon$ represents the fermion loop in the MT diagram. It is worth noticing that the typical scale of $\varepsilon\sim T$, whereas $\{Dq^2,\omega\}\sim T-T_c$. For this reason, the above expression can be simplified by setting the hyperbolic cosine in the denominator to unity and integrating the product of two cooperons $\int d\varepsilon C^R(2\varepsilon+\omega,q) C^A(2\varepsilon+\omega,q)=(\pi/2)C(q)$, where we define the static cooperon $C(q)=[Dq^2+\tau^{-1}_{\phi}]^{-1}$. The resulting variance of the anomalous MT diagram then reads 
\begin{align}
&\langle(\delta\sigma^{\mathrm{an}}_{xx})^2\rangle=\left(\frac{e^2\nu D}{4T}\right)^2\sum_{q_1q_2}C(q_1)C(q_2)\nonumber \\ 
&\int d\omega_1d\omega_2\coth(\omega_1/2T)\coth(\omega_2/2T)M_{12}(\omega,q).
\end{align}
In the most interesting regime of weak dephasing, $\tau_\phi\gg\tau_{\mathrm{GL}}$, where $\tau_{\mathrm{GL}}=\pi/8(T-T_c)$ is the Ginzburg-Landau time, one finds after integrations 
\begin{equation}\label{eq:rms-sigma-MT}
\rms \sigma^{\mathrm{an}}_{xx}\simeq\sigma_Q\mathrm{Gi}\frac{L_T}{L}\left(\frac{T_c}{T-T_c}\right)^2\ln(\tau_\phi/\tau_{\mathrm{GL}}),
\end{equation}
so it is similar to Eq. \eqref{eq:rms-sigma-xx} with an extra logarithmic factor. In the opposite regime of strong dephasing, $\tau_\phi\ll \tau_{\mathrm{GL}}$, the MT term is further suppressed: 
\begin{equation}
\rms \sigma^{\mathrm{an}}_{xx}\simeq\sigma_Q\mathrm{Gi}\frac{L_T}{L}(T_c\tau_\phi)\left(\frac{T_c}{T-T_c}\right). 
\end{equation}

\subsection{Mesoscopic Nernst effect and susceptibility fluctuations}

We can build on this result to consider emergent mesoscopic fluctuations in the transverse thermoelectric coefficient. We start from Eq.~\ref{eq:K-xy} where we need only contributions linear in $Q$, which can be easily extracted by expanding the pair propagator and noticing that 
\begin{equation}
\frac{\partial P(\omega,q)}{\partial q_x}=-B_x(q)P^2(\omega,q).
\end{equation} 
As a next step, we have to sum the resulting expression for $K^{eh}_{xy}(\Omega_m,Q)$ in Eq. \eqref{eq:K-xy} over the Matsubara frequency, as in the case of the conductivity calculation, by contour integration in the complex plane with the help of Eq. \eqref{eq:f}. Completing these steps we arrive at 
\begin{align}
&\beta_{xy}=\frac{4e^2H}{c\pi
T}\sum_{q}B^2_x(q)B^2_y(q)\int d\omega\coth(\omega/2T)\nonumber\\
&\left\{[\Re P^R(\omega,q)]^3\Im P^R(\omega,q)
+\Re P^R(\omega,q)[\Im P^R(\omega,q)]^3\right\}.
\end{align}
Now using Eq. \eqref{eq:P-B-xy} and performing frequency and momentum integrations we find $\beta_{xy}=(e/2\pi)(\xi_{\mathrm{GL}}/l_H)^2\propto (T-T_c)^{-1}$, where $l_H=\sqrt{c/eH}$ is the magnetic length. One should notice that $\beta_{xy}$ has the same scaling with temperature as the conductivity $\sigma_{xx}$. As shown by Ussishkin \cite{N-Th-Ussishkin-2} the magnetization contribution has the same structural form but comes with the coefficient $-1/3$ instead of $1/2$ so that $\alpha_{xy}=\beta_{xy}+cM_z/T$ has an overall coefficient of $1/6$. To address the mesoscopic part of $\beta_{xy}$ we take its variation, square the result, and average over the disorder realization with the help of the correlation function Eq. \eqref{eq:H}. In doing so we encounter quite a cumbersome expression with several contributions to $\langle\delta\beta^2_{xy}\rangle$, but we make an observation that all the emergent terms have exactly the same scaling with temperature, and dependence on the system size, and differ from each other only by a numerical coefficient of the order of unity. For brevity we present here one particular such term 
\begin{align}
&\langle\delta\beta^{2}_{xy}\rangle=A\left(\frac{e^2\nu HL_T}{cgTL}\right)^2
\sum_{q_1q_2}B^2_x(q_1)B^2_x(q_2)B^2_y(q_1)B^2_y(q_2)\nonumber
\\ &\int d\omega_1d\omega_2\coth(\omega_1/2T)\coth(\omega_2/2T)
\nonumber \\ 
&[\Re P^R(\omega_1,q_1)]^4[\Re P^R(\omega_2,q_2)]^4
\Im P^R(\omega_1,q_1)\Im P^R(\omega_2,q_2)
\end{align}
and carry out the remaining calculation up to a factor modulo one (we will absorb all the numerical factors into the redefinition of coefficient $A$). Since most relevant frequencies $\omega\sim T-T_c$ are small compared to temperature we can approximate $\coth(\omega/2T)\approx 2T/\omega$. Transforming the above expression into dimensionless variables
\begin{align}
&\langle\delta\beta^{2}_{xy}\rangle=\frac{e^2A}{g^2}
\left(\frac{\xi^2
L_T}{\ell^2_HL}\right)^2\nonumber \int^{2\pi}_{0}d\phi_1d\phi_2\sin^22\phi_1\sin^22\phi_2
\\ 
&\int^{\infty}_{0}\int^{+\infty}_{-\infty}
\frac{x^2_1x^2_2(x_1+\epsilon)^4(x_2+\epsilon)^4dx_1dx_2dy_1dy_2}{[(x_1+\epsilon)^2+y^2_1]^5[(x_2+\epsilon)^2+y^2_2]^5},
\end{align}
followed by rescaling and integration, one finds
\begin{equation}\label{eq:rms-beta-xy}
\rms\beta_{xy}=\beta_Q\mathrm{Gi}\left(\frac{\xi_{\mathrm{GL}}}{l_H}\right)^2\left(\frac{L_T}{L}\right)\left(\frac{T_c}{T-T_c}\right),
\end{equation}
where we introduced a quantum unit of thermopower $\beta_Q=e/(2\pi\hbar)$. It remains to consider fluctuation-induced corrections to magnetic susceptibility and its mesoscopic fluctuations. From Eq.~\ref{eq:chi} we get for the Aslamazov-Larkin term
\begin{equation}
\chi=-\frac{16e^2}{3c^2}T\sum_{\omega_m,q}\Pi'_{x}P^3(\omega_m,q)\left[\Pi'_{x}\Pi''_{yy}-\Pi'_{y}\Pi''_{xy}\right],
\end{equation}
where derivatives of the polarization operator can be easily computed: $\Pi'_{x,y}=-(\pi\nu D/4T)q_{x,y}$, $\Pi''_{yy}=-(\pi\nu D/4T)$, and $\Pi''_{xy}=0$. Already at this level, by simple power counting of integration variables, one can deduce that $\chi\propto T_c/(T-T_c)$. Consequently one expects that $\langle\delta\chi^2\rangle$ will also scale with $T-T_c$ in the same way as the conductivity and thermomagnetic coefficients. Indeed, 
\begin{align}
&\langle\delta\chi^2\rangle=A\left(\frac{e^2\nu^2\eta L_T}{c^2gL}\right)^2\sum_{q_1q_2}B^2_x(q_1)B^2_x(q_2)\nonumber \\ &\int d\omega_1d\omega_2\coth(\omega_1/2T)\coth(\omega_2/2T)\nonumber \\ 
&\Im[P^R(\omega_1,q_1)]^4\Im[P^R(\omega_2,q_2)]^4,
\end{align}
which, as in the previous examples, reduces with standard steps to
\begin{equation}\label{eq:rms-chi}
\rms\chi=\chi_P\mathrm{Gi}\left(\frac{L_T}{L}\right)\left(\frac{T_c}{T-T_c}\right)^2,
\end{equation} 
where $\chi_P$ is the Pauli susceptibility in the diffusive metal. 

\subsection{Mesoscopic fluctuations of thermopower}

We briefly discuss next the longitudinal thermopower $\alpha_{xx}$ (Seebeck coefficient) and transverse conductivity $\sigma_{xy}$ (Hall coefficient). The Aslamazov-Larkin contribution to $\alpha_{xx}$ is found from the mixed electric-heat currents Kubo response function
\begin{align}
&\alpha_{xx}=-\frac{1}{T}\lim_{\Omega\to0}\frac{1}{\Omega}\mathrm{Im}[K^{eh}_{xx}(\Omega)]^R,\nonumber\\
&K^{eh}_{xx}(\Omega_\nu)=2ieT\sum_{q\omega}\omega_n
B^2_{x}(q)P(\omega_n,q)P(\omega_n+\Omega_\nu,q).
\end{align}
Summation over the Matsubara frequency $\omega_n$ and analytical
continuation follows the same way as in the case of the conductivity calculation, and we obtain
\begin{equation}
\alpha_{xx}=\frac{e}{2\pi
T^2}\sum_q B^2_x(q)\int\frac{\omega
d\omega}{\sinh^2(\omega/2T)}[\Im P^R(\omega,q)]^2\,.
\end{equation}
Without particle-hole asymmetry, $\alpha_{xx}$ is zero. Indeed, $[\Im P^R]^2$ is even in frequency while the rest of the integrand is odd. We have to use a generalized form of the pair propagator. 
Gauge invariance dictates that \cite{Aronov}
\begin{equation}
P(\omega_m,q)=-\frac{1}{\nu}\frac{1}{\pi D
q^2/8T+\epsilon+\pi|\omega_m|/8T+\Upsilon_\omega},
\end{equation}
which generalizes Eq. \eqref{eq:P} to include explicitly the particle-hole asymmetry factor $\Upsilon_\omega=(\mathrm{i}\omega_m/2T_c)(\partial
T_c/\partial E_F)$ that accounts for the gradient of $T_c$ at the Fermi surface. Expanding $P^R$ to the leading linear in $\Upsilon_\omega$ order produces
\begin{equation}
\alpha_{xx}=-\frac{e\nu}{\pi
T^2}\sum_q B^2_x(q)\int
\frac{\omega\Upsilon_\omega\Im P^R(q,\omega)\Im[P^R(\omega,q)]^2 d\omega}{\sinh^2(\omega/2T)},
\end{equation}
where now both propagators are taken at $\Upsilon_\omega\to0$. The resulting expression for $\alpha_{xx}$ is logarithmically divergent in momentum $x$-integration that has to be regularized by introducing an upper cut-off $x_{\mathrm{max}}\simeq1/\epsilon$ [in the original notations this corresponds to $(\xi_{\mathrm{GL}}q_{\mathrm{max}})^2\simeq1$]. This choice is natural since $P^R$, in the form we use here, was obtained from the expansion in small momenta, which is justified only as long as $\mathrm{max}\{Dq^2,\omega\}< T$. As a consequence, 
$\alpha_{xx}\simeq\beta_Q(T/E_F)\gamma_{\mathrm{pha}}\ln[T_c/(T-T_c)]$, where $\gamma_{\mathrm{pha}}=(d\ln T_c/d\ln
E_F)$. Extending this analysis to account for mesoscopic fluctuations gives as a final result 
\begin{equation}
\rms \alpha_{xx}\simeq \beta_Q(T/E_F)\gamma_{\mathrm{pha}}\mathrm{Gi}\left(\frac{L_T}{L}\right)\left(\frac{T_c}{T-T_c}\right).     
\end{equation}
Generally speaking, in a disordered sample, mesoscopic fluctuations of thermopower are present and strong even in the absence of superconducting fluctuations, since per our earlier discussion in the introductory section, particle-hole asymmetry is broken from the scale of Fermi energy down to Thouless energy.    

\subsection{Mesoscopic Hall effect fluctuations}

In order to calculate the Hall coefficient, we need to know the transverse component of the current-current correlation function
$K^{ee}_{xy}$. In the presence of Landau quantization the vertex in real space becomes an operator
$\hat{B}_{i}=-2\nu\eta(-\mathrm{i}\nabla_i+2eA_i)$, where we choose the vector potential in the Landau gauge $A=(0,Hx,0)$. Different components of the vertex, $\hat{B}_x$ and $\hat{B}_y$, do not commute and the matrix elements are
\begin{equation}
\hat{B}^{nn'}_{i}=-\frac{2\sqrt{2}\nu\eta}{l_H}\left\{
\begin{array}{ll}
\mathrm{i}\langle n|\hat{a}-\hat{a}^\dag|n'\rangle & i=x \\
\langle n|\hat{a}+\hat{a}^\dag|n'\rangle & i=y
\end{array}\right.,
\end{equation}
where $\hat{a},\hat{a}^\dag$ are the harmonic oscillator operators. Recalling that $\langle n|\hat{a}|n'\rangle=\langle
n'|\hat{a}^\dag|n\rangle=\sqrt{n}\delta_{n,n'+1}$, we see that only transitions between nearest Landau levels $n\to n\pm1$ are allowed. With these ingredients one finds for the Matsubara response kernel
\begin{align}
&K^{ee}_{xy}(\Omega)=\frac{(4e\nu\eta)^2}{8\pi l^4_H}
T\sum_\omega\sum^{\infty}_{n=0}
(n-1)\nonumber \\ 
&[P_{n+1}(\omega,q)P_n(\omega-\Omega,q)-P_n(\omega,q)P_{n+1}(\omega-\Omega,q)]\,.
\end{align}
After an analytic continuation one gets 
\begin{align}
&\sigma_{xy}=-\frac{(4e\nu\eta)^2}{4\pi^2l^4_H}\sum^{\infty}_{n=0}(n+1)\int d\omega
\coth(\omega/2T)\nonumber\\ 
&\left[\Im P^R_n(\omega,q)\partial_\omega\Re P^R_{n+1}(\omega,q)-
\Im P^R_{n+1}(\omega,q)\partial_\omega\Re P^R_{n}(\omega,q)\right].
\end{align}
In the weak field limit, $H\to0$, one needs only the first term in the expansion in powers of $1/n$ and then replace summation over $n$ by an integration: $(1/l_H)^2\sum_n\to\sum_q$. Taking into account $\partial_n P_n=2\nu(\eta/l^2_H)P^2_n$ and, after some algebra the previous expression can be reduced to 
\begin{equation}
\sigma_{xy}=-\frac{(4e\eta)^2\nu^3\eta}{3\pi Tl^2_H}\sum_q
q^2\int\frac{[\Im P^R(\omega,q)]^3d\omega}{\sinh^2(\omega/2T)},
\end{equation}
where we also used integration by parts with respect to the energy variable. Since $[\Im P^R(\omega,q)]^3$ is odd in energy then without particle-hole asymmetry $\sigma_{xy}$ vanishes.
As in the case of thermopower we expand $P^R$ to the lowest nonvanishing order in $\Upsilon_\omega$ and integrate to find $\sigma_{xy}=(e^2/48)(\omega_c\tau_{\mathrm{tr}})
\gamma_{\mathrm{pha}}[T_c/(T-T_c)]^2$, where $\omega_c=eH/mc$ is the cyclotron frequency, and $\tau_{\mathrm{tr}}$ is the transport scattering time. The mesoscopic part of $\sigma_{xy}$ is then given by  
\begin{equation}\label{eq:rms-sigma-xy}
\rms \sigma_{xy}\simeq \sigma_Q(\omega_c\tau_{\mathrm{tr}})\gamma_{\mathrm{pha}}\mathrm{Gi}\left(\frac{L_T}{L}\right)\left(\frac{T_c}{T-T_c}\right)^3.     
\end{equation}
It should be recalled that the transverse conductance $\sigma_{xy}$  of small normal samples at low temperatures displays universal fluctuations similar to those in $\sigma_{xx}$, however unlike $\sigma_{xx}$ these are asymmetric in the magnetic field \cite{Ma-Lee,Chaltikian}. Furthermore, those contributions are small in the limit when normal metal coherence length $L_T$ is much smaller than the sample size $L$, due to thermal smearing effects. As a consequence, in the vicinity of $T_c$ of superconducting samples the effect is dominated by Eq. \eqref{eq:rms-sigma-xy}. 

\section{Discussion}\label{sec:discussion}

The main results of this work are expressions Eqs.~\eqref{eq:rms-sigma-xx}, \eqref{eq:rms-sigma-MT}, \eqref{eq:rms-beta-xy}, and \eqref{eq:rms-chi} for variances of different kinetic coefficients in mesoscopic superconductors. Because of the long-range phase coherence developing close to $T_c$, sample-specific mesoscopic fluctuations should be observable at large length scales. Similarly to normal samples, these fluctuations are sensitive to magnetic field strength, impurity configuration, and gate voltage. However, in sharp contrast to the normal case, where such fluctuations are universal, interaction effects in the Cooper channel trigger a great amplification of fluctuations due to pairing correlations. This interplay of coherent impurity scattering and interactions leads to an interesting example of quantum mesoscopic phenomena occurring at a macroscopic scale. Despite the fact that mesoscopic fluctuations are no longer universal, we have discovered a different kind of universality in the sense of temperature dependence, which was found to have the same power-law scaling for considered kinetic coefficients in the Ginzburg region of fluctuation-induced transport. 

It is instructive to estimate the order of magnitude for these effects. First, we notice that all terms contain an extra smallness in $\mathrm{Gi}\sim 1/g \ll1$, which is natural for a quantum interference correction. However, the resulting variation of fluctuations has strongly pronounced temperature dependence, $\rms\{\sigma_{xx},\beta_{xy},\chi\}\propto \mathrm{Gi}\,[T_c/(T-T_c)]^2$, which is more singular than the corresponding dependence of their mean values $\{\sigma_{xx},\beta_{xy},\chi\}\propto T_c/(T-T_c)$. We observe that for the system size $L\sim\xi_{\mathrm{GL}}$ and at the threshold of applicability of the Gaussian theory of superconducting fluctuations, $T-T_c\sim \mathrm{Gi}\,T_c$, the scale of conductance fluctuations [per Eq. \eqref{eq:rms-sigma-xx}] is of the order of $\rms\sigma_{xx}\sim\sigma_Q\sqrt{g}$, which is parametrically bigger than UCF in normal metals. At the same time, these fluctuations are still smaller than the bare Drude value of the normal state conductivity, $\rms \sigma_{xx}/\sigma\sim \sqrt{\mathrm{Gi}}\ll1$. The situation is different for the transverse magnetic thermopower, because for a particle-hole symmetric case the normal state quasiparticle contribution is absent and we have to compare the mesoscopic part directly to the fluctuation-induced term, so that $\rms \beta_{xy}/\beta_{xy}\sim L_T/L$, where we assume sufficient proximity to $T_c$. Under the same provisions, fluctuations in magnetic susceptibility are as strong as in the conductivity, this conclusion carries over to fluctuations in $\rms \alpha_{xy}$, so that one should expect large reproducible mesoscopic noise of the overall Nernst signal.     

The calculations presented in this work have been carried out for homogeneously disordered superconductors. Therefore, our results cannot be directly compared to the existing experimental findings in which the samples were granular in their origin \cite{Frydman-SSC98,Frydman-PU98,Zadorozhny-PRB01,Dynes-PRB98}. Granularity adds another parameter into the model -- inter-grain conductance -- which leads to a strong competition between Aslamazov-Larkin, Maki-Thompson, and DOS effects \cite{Lerner}. Nonetheless, the main features predicted by the theory should be present for inhomogeneously disordered superconductors as well. Indeed, the predicted sample-specific conductance fluctuations were observed experimentally in samples of macroscopic length, and only in a narrow temperature range in the immediate vicinity of $T_c$, consistent with the theory. The amplitude of the conductance fluctuations was found to greatly exceed that of the UCF in normal samples. It should also be emphasized that some other features accompanying giant mesoscopic effects, such as suppression of $h/2e$ oscillations in cylindrical samples, negative magnetoresistance, and its asymmetry, can be also addressed within the same theoretical model. Currently, we are unaware of experimental measurements of mesoscopic effects in the thermomagnetic transport of superconductors. The mesoscopic Nernst effect has been studied experimentally only in the non-superconducting systems \cite{Goswami-MesoNernstExp}. Verification of the temperature scaling and the overall magnitude of the effect for mesoscopic fluctuations of the Nernst coefficient predicted here would provide an important test for our understanding of thermomagnetic transport phenomena in correlated systems. 

As an outlook, we briefly mention possible extensions of this work geared towards future research in the area of magneto-thermo-transport phenomena in superconductors. The regime of quantum fluctuations is of great interest. One could distinguish regimes where superconductivity is suppressed by orbital or spin effects. In particular, in the latter case of Pauli-limited superconductivity, fluctuation effects are dominated by virtual rather than real pair excitations \cite{Khodas}. The Nernst effect has not been studied for this scenario. It is also of special importance to investigate possible mechanisms for extrinsic skew-scattering and side-jump contributions to the Nernst effect which generally play a crucial role in the anomalous and nonlinear Hall effects. Last, there is enough motivation to attempt computations beyond the perturbation Gaussian theory of superconducting fluctuations by adopting a strong-coupling Eliashberg approach. This analysis of the Nernst effect will be certainly relevant for high-$T_c$ materials.    
\subsection*{Acknowledgment}

We would like to thank Boris Altshuler, Lev Ioffe, Karen Michaeli, and Mikhail Skvortsov for discussions. We are grateful to Kamran Behnia for detailed explanations of the Nernst effect measurements in superconductors per Ref. \cite{Nernst-Review}. This work was supported in part by BSF Grant No. 2016317, ISF Grant No. 1287/15, and NSF Grant No. DMR-1653661.

\end{document}